# Colloidal Directional Structures at a Nematic Liquid Crystal–Air Interface

Nan Wang[1,2,3], Julian Evans[2,*], Chenxi Li[4], Victor M. Pergamenshchik[5,6], Sailing He[1,2,7,*]

*1 Ningbo Research Institute, Zhejiang University, Ningbo 315100, China.*

*2 Centre for Optical and Electromagnetic Research, College of Optical Science and Engineering, National Engineering Research Center for Optical Instruments, Zhejiang University, Hangzhou 310058, China.*

*3 School of Information Science and Engineering, NingboTech University, Ningbo 315100, China.*

*4 Key Laboratory of Spectral Imaging Technology, Xi'an Institute of Optics and Precision Mechanics, Chinese Academy of Sciences, Xi'an 710119, China.*

*5 Institute of Physics, National Academy of Sciences of Ukraine, Kyiv 03039, Ukraine.*

*6 Center for Theoretical Physics, Polish Academy of Sciences, Warsaw 02668, Poland.*

*7 Department of Electromagnetic Engineering, School of Electrical Engineering, Royal Institute of Technology, SE-100 44 Stockholm, Sweden.*

\* julian.evans@colorado.edu, sailing@zju.edu.cn

# Colloidal Directional Structures at a Nematic Liquid Crystal–Air Interface

We present a variety of structures formed by colloidal droplets at a nematic liquid crystal–air interface, where the elastic dipole-dipole, quadrupole-quadrupole, and dipole-quadrupole interactions are all essentially involved. The colloidal structures observed not only include chains with kinks or clusters, but also comprise directional structures, such as directional chains and branches, whose direction is associated with the tilting director in the liquid crystal layer. The dipole-quadrupole interaction, originating from the polydispersity of the droplets, plays a central role for the formation of these directional structures. Clusters consisting of directional branches and chains are also observed and found to be fractal statistically.

**Introduction**

Colloids in nematic liquid crystals (LCs) can self-assemble into a diversity of ordered structures [1,2], ranging from chains [3-6] to various colloidal crystals [5-14]. These structures can be observed either in a nematic LC bulk [3-5,7-10,14,15] or at a nematic LC surface [6,11-13]. Understanding the self-assembly of colloidal objects in nematic LCs will improve our grasp of complex fluids and develop new methods of constructing artificial colloidal structures [1,2]. The unique physics behind such nematic colloids lies in the long range interaction of elastic multipoles which is analogous to, but even richer than the electrostatic multipole interaction [16-23]. Previously we have reported a controlled cascade of self-assembly of colloidal droplets at a nematic LC–air interface, where elastic dipole-dipole interaction is the dominant factor [6]. For thinner LC layers or larger droplets, however, elastic quadrupoles induced by the colloids begin to appear and lead to more complex self-assembled structures. A distinguishing character of such system is the formation of directional structures, such as directional chains or branches, which can most generally describe a broad class of real world materials such as corals [24], snowflakes [25], dendrimers [26,27], and particle aggregates with many branches [28,29]. In addition, directional structures can be a central concern to many chemical and physical processes. Realizing directional structures by nematic colloids is particularly challenging since they have highly symmetric interaction potentials for most cases.

In this paper, we demonstrate a diversity of structures formed by colloidal droplets at a nematic LC–air interface, where the colloids induce both elastic dipoles and quadrupoles. Using a thin film of nematic LCs between a layer of fluid and air induces a director tilt by the hybrid alignment [30]. The observed colloidal structures include not only those which are not tilt connected, but also comprise directional chains and branches, whose direction is determined by the director tilt. Elastic dipole-dipole (D-D), quadrupole-quadrupole (Q-Q), and dipole-quadrupole (D-Q) interactions account for the formation of the various structures. The D-Q interaction is only present for droplets of different sizes and is vital for the formation of the directional structures. When the number of droplets is sufficiently large, colloids can form clusters which consist of branches and chains and at the same time are stabilized by many-body elastocapillary effect [6,31,32]. Such clusters are found to be fractal statistically.

**Materials and methods**

The sample is made by mixing a nematic LC E7 (HCCH, China) with 65 wt % sulfuric acid. The layer of the sulfuric acid is covered by the LC layer, whose thickness $h$ can be set in the range $h \sim 0.9-10 \mu m$ by adopting different amount of LCs. At the LC-air interface, the LC director tends to be perpendicular to the interface due to the homeotropic anchoring; while at the LC-acid interface, the director tends to be tangential to the interface due to the degenerate planar anchoring. The actual boundary tilts of the director are determined by $h$. When the LC layer is thinner, the director at the LC-air interface deviates more from the interface normal. Numerous droplets are generated as soon as the LC contacts the acid, and a substantial amount of them are trapped at the LC-air interface. The droplets coalesce and grow for minutes due to Brownian motion and LC flowing. The coalescence finally concludes and is prevented by the sufficiently large director distortions due to the anchoring at the LC-droplet interface, when the radius of the droplets reaches some stabilizing size $R$. The droplet size and uniformity are substantially affected by LC flowing, which is experimentally adopted as a factor to adjust the parameters. When the surface of the sample is made flat by reducing the contact angle of the acid solution on the glass, LC flowing is weak; otherwise, the surface of the sample has tilt and LC flowing is violent.

**Results and discussions**

The effective director tilt $\Psi$ at the droplet center is determined by the droplet size $R$ and thickness $h$. $\Psi$ is larger for larger $R$ and smaller $h$. Three typical configurations of isolated droplets at the LC-air interface are observed. In thick LC films ($h \sim 10 \mu m$), the LC director $\hat{\mathbf{n}}$ at the LC-air interface is vertical, and one point defect, boojum, is produced near the lower pole of the droplet due to the tangential anchoring at the LC-droplet interface (Fig. 1a). The droplet induces an uniaxial elastic dipole $d$. In this case, the horizontal projection of $\hat{\mathbf{n}}$ near the droplet is radial in the $x_1x_2$ plane (Inset of Fig. 1b). This is consistent with the interference color revealed in the polarized optical microscopic (POM) images with a full-wave plate ($\lambda \sim 530 nm$) of the droplet (Fig. 1b). The interference color remains generally unchanged when the sample is rotated around the $x_3$ axis by 45° (Fig. 1c). In thin LC films ($h \sim 5\mu m$), as $\hat{n}$ deviates from the surface normal at the LC-air interface, the boojum is shifted from the lower pole of the droplet, and the director at the droplet center shows greater tilt $\Psi$ (Fig. 1d). The elastic dipole induced by the droplet contains both standard uniaxial $d$ dipole and negative $\gamma$ or banana dipole [6,19]. We use vector $\vec{n}_\parallel$ to represent the tilting direction of the uniaxial dipole $d$, which is also related to the director bending in the $x_1 x_3$ plane. Inset of Fig. 1e demonstrates the horizontal projection of $\hat{n}$ in the $x_1 x_2$ plane and this pattern is verified by the interference color of the POM image (Fig. 1e). When the sample is rotated around the $x_3$ axis by 45°, the interference color changes and generally indicates that the average direction of the horizontal component of $\hat{n}$ is along the $x_1$ axis (Fig. 1f). As the film becomes thinner ($h \sim 3\mu m$), another boojum appears on the other side of the droplet (Fig. 1g). The droplet induces a pair of asymmetric uniaxial $d$ dipoles and a negative $\gamma$ dipole. The two uniaxial $d$ dipoles with opposite signs result in a quadrupole $Q$ and a net $d$ dipole. The horizontal projection of $\hat{n}$ in the $x_1 x_2$ plane is tangential to the droplets at its boundary (inset of Fig. 1h). As a result, the POM image of Fig. 1i shows a different pattern from Fig. 1b -- yellow and blue switches their positions. The different brightness of the interference color in Fig. 1i reflects the asymmetric elastic dipoles. The axis connecting boojums could be identified along the $x_1$ axis by the POM image of the sample rotated around the $x_3$ axis by 45° (Fig. 1j).

In reference [6], we have reported the self-assembling of the colloidal droplets mainly associated with elastic dipoles, here we focus on the phenomena which also involves elastic quadrupoles. [16]The droplets at the LC-air interface generally induce both elastic dipoles and elastic quadrupoles when the size of the droplets is comparable to the thickness of the LC layer. In the reference frame where $x'_3$ axis coincides with the effective director tilt $\Psi$ at the droplet center (Fig. 2a), multipoles can be considered as consisting of two components related to the two transverse coordinates x'$_1$ and x'$_2$. The two components of elastic dipoles induced by the droplet have the form [16-19,23]

$$\vec{D}_1 = (d, 0, -\gamma) \\ \vec{D}_2 = (0, d, 0),\qquad(1)$$

and those of the elastic quadrupole are

$$\ddot{Q}_1 = Q \begin{pmatrix} 0 & 0 & -1 \\ 0 & 0 & 0 \\ -1 & 0 & 0 \end{pmatrix} \qquad \ddot{Q}_2 = Q \begin{pmatrix} 0 & 0 & 0 \\ 0 & 0 & -1 \\ 0 & -1 & 0 \end{pmatrix},\qquad(2)$$

where $d > 0$, $\gamma > 0$ and $Q > 0$. The strength of the net $d$ dipole, $\gamma$ dipole, and the quadrupole $Q$ depend on $\Psi$. When $\Psi$ increases, $d$ weakens whereas $\gamma$ and $Q$ strengthen.

The elastic interaction potential $U$ between two droplets contains three terms, $U = U_{D-D} + U_{D-Q} + U_{Q-Q}$, the potential of D-D ($U_{D-D}$), D-Q ($U_{D-Q}$) and Q-Q ($U_{Q-Q}$) interactions. Let $\vec{r}$ be the separation vector of two droplets (indicated by the superscripts I and II) shown in Fig. 2b, then $U_{D-D}$, $U_{D-Q}$, and $U_{Q-Q}$ are

$$U_{D-D} = \sum_{t=1,2} -\frac{12\pi K}{r^3}[\vec{D}_t^I \cdot \vec{D}_t^{II} - 3(\vec{D}_t^I \cdot \vec{u})(\vec{D}_t^{II} \cdot \vec{u})]\qquad(3)$$

$$U_{D-Q} = \sum_{t=1,2} -\frac{16\pi K}{r^4}[5\ddot{Q}_t^I : (\vec{u}\vec{u})(\vec{D}_t^{II} \cdot \vec{u}) - 2\ddot{Q}_t^I : (\vec{D}_t^{II}\vec{u}) - 5\ddot{Q}_t^{II} : (\vec{u}\vec{u})(\vec{D}_t^I \cdot \vec{u}) + 2\ddot{Q}_t^{II} : (\vec{D}_t^I\vec{u})]\qquad(4)$$

$$U_{Q-Q} = \sum_{t=1,2} -\frac{20\pi K}{3r^5}[35\ddot{Q}_t^I : (\vec{u}\vec{u})\ddot{Q}_t^{II} : (\vec{u}\vec{u}) - 20(\ddot{Q}_t^I \cdot \vec{u}) \cdot (\ddot{Q}_t^{II} \cdot \vec{u}) + 2\ddot{Q}_t^I : \ddot{Q}_t^{II}],\qquad(5)$$

where $K$ is the elastic constant and $\vec{u} = \frac{\vec{r}}{r} = (-\cos\varphi\cos\psi, -\sin\varphi, -\cos\varphi\sin\psi)$.

Substituting equations (1, 2) into equation (3-5), we obtain

$$U_{D-D} = -\frac{12\pi K}{r^3}[(3\cos^2\varphi\sin^2\psi - 1)d^I d^{II} - 3\cos^2\varphi\sin\psi\cos\psi(d^I\gamma^{II} + d^{II}\gamma^I) \\ + (1 - 3\cos^2\varphi\sin^2\psi)\gamma^I\gamma^{II}]\qquad(6)$$

$$U_{D-Q} = \frac{32\pi K}{r^4}[(Q^I d^{II} - Q^{II} d^I)\cos\varphi\sin\psi(3 - 5\cos^2\varphi\sin^2\psi) \\ + (Q^I\gamma^{II} - Q^{II}\gamma^I)\cos\varphi\cos\psi(5\cos^2\varphi\sin^2\psi - 1)]\qquad(7)$$

$$U_{Q-Q} = \frac{80\pi K}{3r^5} Q^I Q^{II} (35\cos^4 \varphi \sin^4 \psi - 30\cos^2 \varphi \sin^2 \psi + 3). \tag{8}$$

Kinked chains formed by droplets with uniform size ($h \sim 3.3 \mu m$, $R \sim 1.3 \mu m$) are observed in our experiment (Fig. 3a). The POM image (Fig. 3b) of the droplets reveals the same pattern of the interference color shown in Fig. 1h, indicating that the induced elastic quadrupoles are prominent. For two identical droplets with equal elastic dipoles and quadrupoles ($d^I = d^{II} = d$, $\gamma^I = \gamma^{II} = \gamma$, $Q^I = Q^{II} = Q$), the potential between them exhibits angular dependence illustrated in Fig. 3c, where for a plausible estimation, the parameters are chosen as $\psi = 65°$, and $d:\gamma:\frac{Q}{r} = 1.5:1:1$. There are four equal minima of the elastic potential located around one droplet which leads to the formation of kinks. When the density of the droplets is sufficiently large, the droplets can also assemble into clusters (Fig. 3a). Ideal colloidal crystals can be formed when all the four minima around each droplet are occupied by other colloids. Such crystals exhibit symmetry in the directions of both $x_1$ axis and $x_2$ axis, even though the director is actually tilted within the LC layer.

With a kinked chain as the backbone, droplets with different radii will attach to it forming a directional chain (Fig. 4a). As shown in Fig. 4a and 4b, smaller droplets attached to the chain will be only located at one specific side of it, so that the chain becomes directional. For example, when we examine two sections (A and B shown in Fig. 4b) belonging to the same kinked chain but pointing to different directions, we find that smaller droplets attached to section A are always on the left side of it, while those attached to section B are always on the right side of it. Compared with the case demonstrated in Fig. 3a, where the director tilting in the LC layer is not manifested by the kinked chain, $\vec{n}_{\parallel}$ can be determined by observing on which side of the directional chain the smaller droplets are located. This is an intriguing difference between a directional chain and an ordinary kinked chain.

In order to explain how the directional chains are formed, we first consider the elastic interaction potential between a pair of droplets with different sizes. In our case, both elastic dipoles and elastic quadrupoles of the two droplets are involved. We estimate the elastic dipoles as $d \propto R^2$, $\gamma \propto R^2$ and the elastic quadrupole as $Q \propto R^3$. Substituting the above relations in equation (7), we obtain that $|U_{D-Q}| \propto |R^I - R^{II}|$, where $R^I$ and $R^{II}$ are the radii of the first and the second droplets. Therefore, the elastic interaction potential between a pair of droplets with different sizes not only contains the components of D-D and Q-Q interactions, but also has the contribution from D-Q interaction, which is not present in the case of two identical droplets. The form of $U_{D-Q}$ indicates that the larger the difference between the sizes of the two droplets, the larger its contribution to the total elastic interaction potential. This term is decisive for the presence of the direction associated with the directional chains.

To specify this issue more clearly, we calculate the elastic interaction potential (denoted by $U_{Droplet-Chain}$) of a smaller droplet and a chain (Fig. 4c). Fig. 4d shows the schematics of this numerical simulation. As a demonstration, we assume that the droplets constituting the chain are identical, and for them the parameters are set as $\psi = 58°$, and $d : \gamma : \frac{Q}{r} = 2:1:1$. For simplicity, we assume that the chain has no kinks. And as a result of the calculation of the elastic interaction potential, the droplets tend to form chains that have a tilting angle $\alpha \approx 30.6°$ (case 1) or $\alpha \approx -30.6°$ (case 2) with $\vec{n}_\parallel$ (Fig. 4d). Then we calculate the total elastic interaction potential of the chain (consisting of $n+1$ droplets) and the smaller droplet

$$U_{Droplet-Chain} = \sum_{i=0}^{n} U_{Droplet-i} = \sum_{i=0}^{n} U_{D-D}^{(i)} + U_{D-Q}^{(i)} + U_{Q-Q}^{(i)},$$

where $U_{Droplet-i}$, the elastic interaction potential of the smaller droplet and the ith droplet of the chain, has three terms $U_{D-D}^{(i)}$, $U_{D-Q}^{(i)}$, and $U_{Q-Q}^{(i)}$ associated with the D-D, D-Q and Q-Q interactions respectively. We set $n=10$, and assume that the radius of the smaller droplet is half of that of the droplets in the chain, so that for the smaller droplet its elastic multipoles roughly satisfy $d' = \frac{1}{4}d$, $\gamma' = \frac{1}{4}\gamma$, and $Q' = \frac{1}{8}Q$. The position of the smaller droplet is described by $\rho$, the distance from the smaller droplet to the chain, and $\delta$, the distance (measured along the direction of the chain) from the center of the 0$^{th}$ droplet to the smaller droplet. Then we calculate the relation between $U_{Droplet-Chain}$ and $\delta$ with a fixed $\rho = 0.75a$ ($a$ is the separation of neighboring droplets in the chain). The smaller droplet can be located on either side of the chain and the corresponding elastic interaction potentials are both calculated. The blue line in Fig. 4c represents $U_{Droplet-Chain}(\delta)$ for the case with the smaller droplet located on the right side of the chain for either $\alpha \approx 30.6°$ or $\alpha \approx -30.6°$ (the side with blue circles in Fig. 4d). The red line in Fig. 4c represents $U_{Droplet-Chain}(\delta)$ for the case with the smaller droplet located on the left side of the chain (the side with red dashed circles in Fig. 4d). And the blue (or red) circles in Fig. 4d label the positions which correspond to each local minimum of $U_{Droplet-Chain}(\delta)$. It is seen in Fig. 4c that the minima associated with the blue line are lower than those of the red line, therefore the positions represented by the blue circles in Fig. 4d are more favorable to hold the smaller droplet. This result is in good agreement with the shapes of sections A and B in the (rotated) experimental picture (Fig. 4d). In addition, it should be noted that if $U_{D-Q}^{(i)} = 0$, the minima associated with the blue line and the red line in Fig. 4c will be equal, indicating that the D-Q interaction is indispensable for the formation of directional chains.

When $h/R$ is relatively large ($h \sim 1.5\mu m$, $R \sim 0.3\mu m$), droplets will form chains originating mainly from the D-D interaction [6]. If the size of the droplets shows polydispersity, two such chains can form a Y-branch when they meet at a larger droplet (Fig. 5a, b). Such branch structure, composed of one large droplet in the center and three small droplets around it, is another typical structure originating from the D-Q interaction. Generally, the large droplet induces a considerable quadrupole component while the small droplets mainly induce dipole components. As a demonstration, we calculate the elastic interaction potential of two droplets with $\psi = 58°$, $d:\gamma:\frac{Q}{r}=2:1:1$ for the large droplet and $d'=\frac{1}{4}d$, $\gamma'=\frac{1}{4}\gamma$, and $Q'=\frac{1}{8}Q$ for the small droplet. The angular dependence of $U$ is shown in Fig. 5c demonstrating that around one large droplet there are three locations with the minimum potentials for a smaller droplet. This result leads to the formation of the Y-branch, whose direction is associated with the director tilting in the LC layer (Fig. 5c). When the density of the droplets is sufficiently large, the droplets assemble into clusters stabilized by the many-body elastocapillary effect that we reported before [6] and generally shows round contour (Fig. 5a). The inner structure of the clusters consists of "trees" formed by the chains and the branches (Fig. 5b). The trees exhibit a power law indicating that they are fractal statistically. For example, when we examine an encircled tree originating from a certain droplet $O$ (Fig. 5b), we divide the tree into several areas by arcs whose center is $O$ and radius is $L$, and use $N(L)$ to denote the number of droplets in the fan encircled by the arc with radius $L$. Then $\ln(N)$ is found to be linearly dependent on $\ln(L)$ (Fig. 5d), which is equivalent to $N(L) \propto L^{\mathcal{D}}$. And thus $\mathcal{D}$ is the fractal dimension of the examined tree. $\mathcal{D}$ is calculated as $\mathcal{D}=1.64\pm0.02$ for the encircled area in Fig. 5b. Applying the same calculation to other trees of the same sample results in good linearity of $\ln(N)$ and $\ln(L)$ as well, and the calculated $\mathcal{D}$ is generally within the range of $\mathcal{D} \sim 1.5-1.8$. The fractal dimension is dependent on a number of factors. One of them might be the number ratio of the large droplets and the small droplets, and this ratio is substantially affected by the LC flowing when the sample is prepared.

**Conclusions**

The colloidal structures originated from the elastic D-D, Q-Q, and D-Q interactions of the droplets at a nematic LC–air interface comprise kinked chains or clusters and a variety of directional structures. These directional structures, including directional chains and branches, result from the asymmetry of the D-Q interaction potential, which is only present for droplets with different sizes. The direction of such structures is associated with the director tilting in the LC layer. In other words, the tilting direction of the director which is originally invisible can be visualized by the directional structures in the macroscale. Another intriguing property of the colloidal structures is that the clusters consisting of directional branches exhibit fractal features.

Acknowledgements

This work was supported by the National Natural Science Foundation of China under Grant 12004332, 62050410447; National Key Research and Development Program of China under Grant 2018YFB2200202; Postdoctoral Science Foundation of China under Grant 2019M662018; VMP's research is part of the project No. 2022/45/P/ST3/04237 co-funded by the National Science Centre and the European Union Framework Program for Research and Innovation Horizon 2020 under the Marie Skłodowska-Curie grant agreement No. 945339.

Declaration of interest statement

The authors report there are no competing interests to declare.

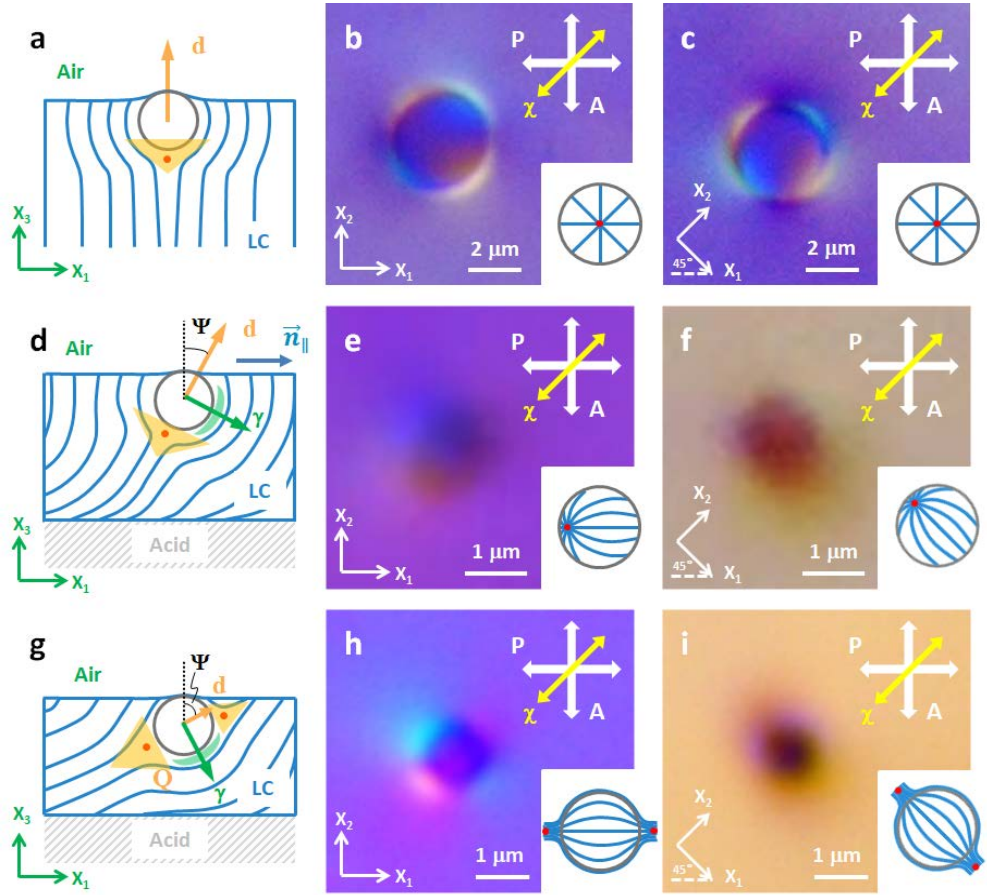

Figure 1. (a) Schematic and (b, c) POM images with a full-wave plate of a droplet inducing a boojum near its lower pole in a thick LC film. (d) Schematic and (e, f) POM images with a full-wave plate of a droplet inducing a boojum shifted to the side in thinner LC film. (g) Schematic and (h, i) POM images with a full-wave plate of a droplet inducing both boojums in the thinnest LC film. (c, f, i) are the images of the droplets rotated around $x_3$ axis clockwise by 45º from their initial positions (b, e, h) respectively. (b, c, e, f, h, i) P and A indicate the polarizations of polarizer and analyzer. χ indicates the slow axis of the full-wave plate. The insets are the schematics in $x_1 x_2$ plane. Red dots in the schematics represent the point defect boojum.

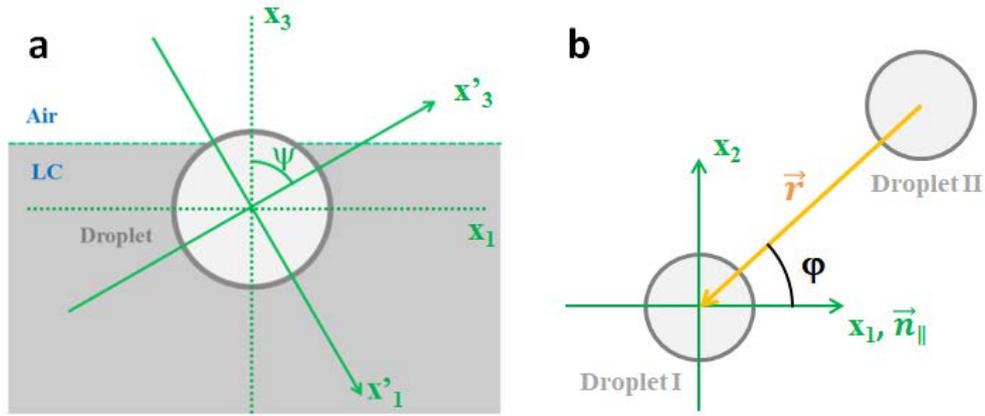

Figure 2. Schematics of (a) one droplet and (b) two droplets at the nematic LC–air interface.

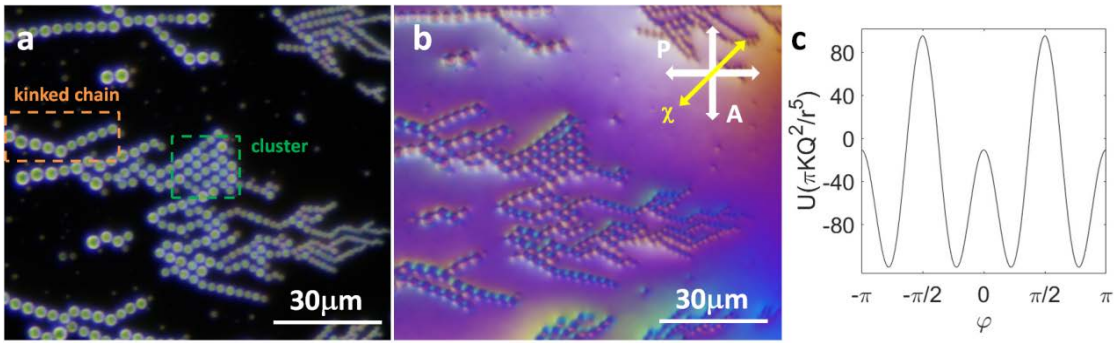

Figure 3. (a) Dark field microscopic image and (b) POM image with a full-wave plate of kinked chains and clusters formed by colloidal droplets at the nematic LC–air interface. P and A indicate the polarizations of polarizer and analyzer. χ indicates the slow axis of the full-wave plate. (c) The elastic interaction potential of two identical droplets with parameters $\psi = 65°$, and $d : \gamma : \dfrac{Q}{r} = 1.5 : 1 : 1$.

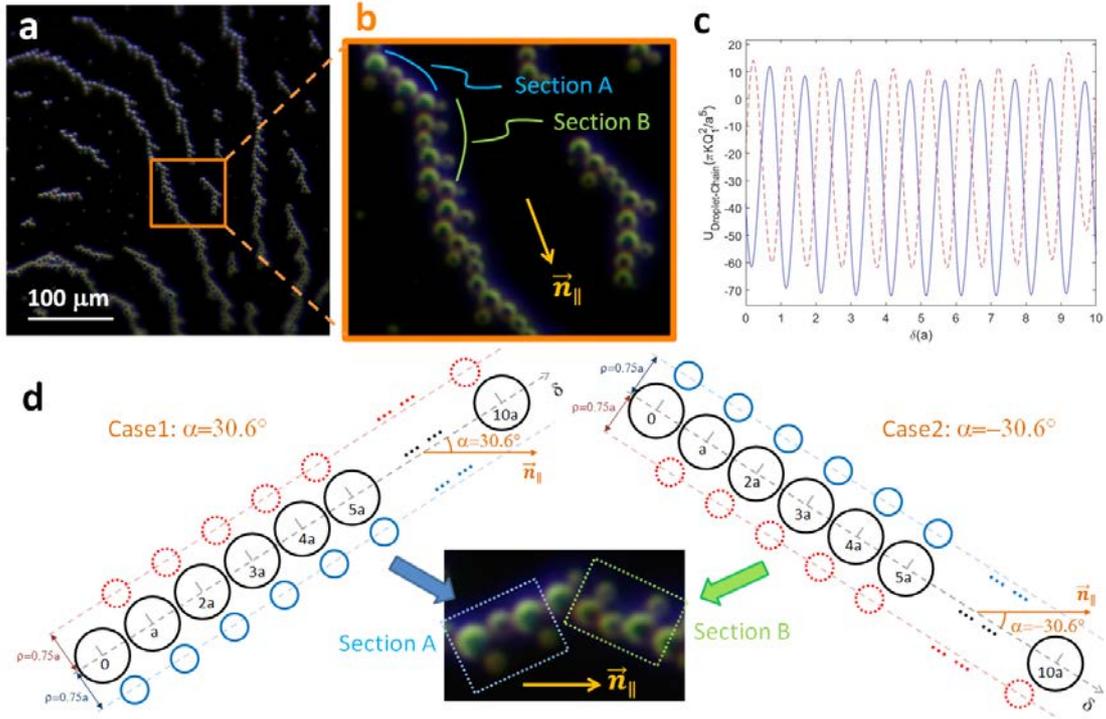

Figure 4. (a) Dark field microscopic image and (b) the zoomed-in image of directional chains formed by polydisperse droplets at the nematic LC–air interface. (c) The elastic interaction potential between a chain and a smaller droplet. (d) Schematics of directional chains. The chain represented by the black circles acting as the backbone can point to two possible directions (case 1: $\alpha \approx 30.6°$ or case 2: $\alpha \approx -30.6°$). The blue line in (c) represents $U_{Droplet-Chain}(\delta)$ for the case with the smaller droplet located on the right side of the chain (the side with blue circles). The red line in (c) represents $U_{Droplet-Chain}(\delta)$ for the case with the smaller droplet located on the left side of the chain (the side with red dashed circles). And the blue (or red) circles in Fig. 4d label the positions which correspond to each local minimum of $U_{Droplet-Chain}(\delta)$. For either case 1 or case 2, the positions represented by the blue circles are more favorable to hold smaller droplets than those represented by the red dashed circles.

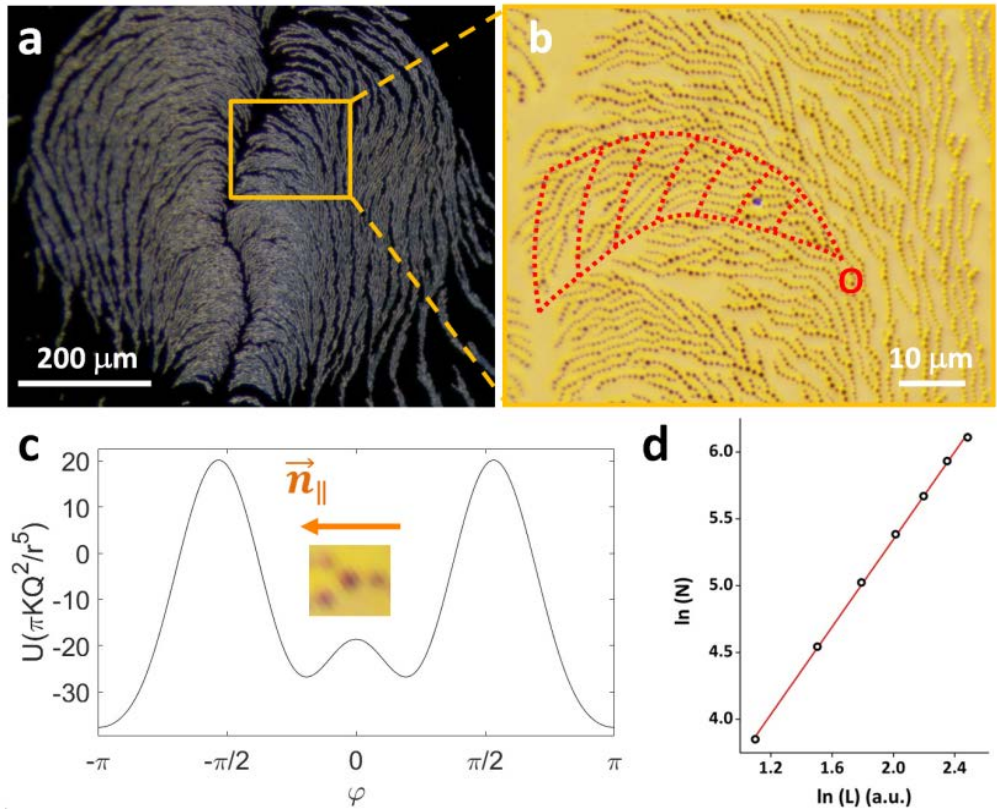

Figure 5. (a) Dark field microscopic image of a cluster consisting of Y-branches and chains of colloidal droplets at the nematic LC–air interface. (b) The zoomed-in bright field image of the inner structure of the cluster. (c) The elastic interaction potential of two droplets with different sizes accounts for the formation of the Y-branch, where around one large droplet there are three smaller droplets. (d) Linear dependence of $\ln(N)$ on $\ln(L)$ gives rise to a fractal dimension $\mathcal{D} = 1.64 \pm 0.02$ for the encircled area in (b).